\documentclass[justified]{tufte-handout}

\title{Introduction to Automatic Backward Filtering Forward Guiding\thanks{}}

\author[Frank van der Meulen]{Frank van der Meulen \\ Vrije Universiteit Amsterdam \\ The Netherlands)}

\usepackage{graphicx} 
  \graphicspath{{graphics/}} 
\usepackage{amsmath}  
\usepackage{booktabs} 
\usepackage{units}    
\usepackage{multicol} 
\usepackage{lipsum}   
\usepackage{fancyvrb} 
  \fvset{fontsize=\normalsize}

\usepackage{latexsym, amsthm, amsfonts,bm,amssymb,mathscinet} 

\usepackage{cancel}

\newcommand{\veryshortarrow}[1][3pt]{\mathrel{%
   \hbox{\rule[\dimexpr\fontdimen22\textfont2-.2pt\relax]{#1}{.4pt}}%
   \mkern-4mu\hbox{\usefont{U}{lasy}{m}{n}\symbol{41}}}}
\newcommand{\pf}{\veryshortarrow}

\usepackage{tikz-cd}
\usetikzlibrary{circuits.logic.US}
\usetikzlibrary{positioning}
\usetikzlibrary{fit}
\usetikzlibrary{cd}
\usetikzlibrary{arrows}
\usetikzlibrary{calc}
\usetikzlibrary{decorations.markings}
\usetikzlibrary{shapes.geometric}
\tikzset{ed/.style={auto,inner sep=2pt,font=\scriptsize}} %
\tikzset{>=stealth}

\tikzset{vert/.style={draw,circle, minimum size=6mm, inner sep=0pt, fill=white}}
\tikzset{vertblank/.style={ minimum size=6mm, inner sep=0pt, fill=white}}

\tikzset{vertbig/.style={draw,circle, minimum size=8mm, inner sep=0pt, fill=white}}
\tikzset{->-/.style={decoration={
      markings,
      mark=at position #1 with {\arrow{>}}},postaction={decorate}}}

\tikzset{edge/.style={line width=0.5pt, decoration={markings,mark=at position 1 with %
    {\arrow[scale=1.5,>=stealth]{>}}},postaction={decorate}}}

\tikzset{dotted/.style={black!30, line width=0.5pt}}

\pgfdeclarelayer{edgelayer}
\pgfdeclarelayer{nodelayer}
\pgfsetlayers{edgelayer,nodelayer,main}

\tikzstyle{morphism}=[fill=white, draw=black, shape=rectangle]
\tikzstyle{medium box}=[fill=white, draw=black, shape=rectangle, minimum width=0.8cm, minimum height=0.9cm]
\tikzstyle{large morphism}=[fill=white, draw=black, shape=rectangle, minimum width=1.7cm, minimum height=1cm]
\tikzstyle{bn}=[fill=black, draw=black, shape=circle, inner sep=1.5pt]
\tikzstyle{effect}=[fill=white, draw=black, regular polygon, regular polygon sides=3, minimum width=0.4cm, inner sep=0pt]
\tikzstyle{state}=[fill=white, draw=black, regular polygon, regular polygon sides=3, minimum width=0.4cm, shape border rotate=180, inner sep=0pt]
\tikzstyle{medium state}=[fill=white, draw=black, regular polygon, regular polygon sides=3, minimum width=1.3cm, inner sep=0pt, shape border rotate=180]
\tikzstyle{large state}=[fill=white, draw=black, regular polygon, regular polygon sides=3, minimum width=2.2cm, shape border rotate=180, inner sep=0pt]
\tikzstyle{wn}=[fill=white, draw=black, shape=circle, inner sep=1.5pt]

\tikzstyle{arrow}=[->]
\tikzstyle{dashed box}=[-, dashed]

\tikzset{none/.style={%
     append after command={%
       \pgfextra{\node [right] at (\tikzlastnode.mid east) {{\tiny\tikzlastnode}};}
     }}}
\tikzstyle{none}=[]

\DeclareSymbolFont{bbsymbol}{U}{bbold}{m}{n}
\DeclareMathSymbol{\bbsemi}{\mathbin}{bbsymbol}{"3B}
\DeclareMathSymbol{\bbcomma}{\mathbin}{bbsymbol}{"2C}

\usepackage{newunicodechar}
\newunicodechar{ä}{\"a}
\newunicodechar{ü}{\"u}
\newunicodechar{ö}{\"o}
\newunicodechar{Ä}{\"A}
\newunicodechar{Ü}{\"U}
\newunicodechar{Ö}{\"O}
\newunicodechar{ß}{\ss}
\newunicodechar{λ}{\ensuremath{\lambda}}
\newunicodechar{θ}{\ensuremath{\theta}}
\newunicodechar{σ}{\ensuremath{\sigma}}
\newunicodechar{ν}{\ensuremath{\nu}}
\newunicodechar{μ}{\ensuremath{\mu}}
\newunicodechar{Σ}{\ensuremath{\Sigma}}
\newunicodechar{ˣ}{\ensuremath{^*}}
\newunicodechar{β}{\ensuremath{\beta}}
\newunicodechar{′}{\ensuremath{^\prime}}
\newunicodechar{∘}{\ensuremath{\circ}}
\newunicodechar{Δ}{\ensuremath{\Delta}}
\newunicodechar{ᵒ}{\ensuremath{^\circ}} 
\newunicodechar{Φ}{\ensuremath{\Phi}} 

\newcommand{\dd}{{\,\mathrm d}}

\renewcommand{\th}{\theta}

\renewcommand{\phi}{\varphi}

\renewcommand{\scr}[1]{{\mathcal #1}}

\newcommand{\argmax}{\operatornamewithlimits{argmax}}
\newcommand{\EE}{\mathbb{E}}

\newcommand{\PP}{\mathbb{P}}
\newcommand{\ind}{\mathbf{1}}

\newcommand{\bem}{\begin{bmatrix}}
\newcommand{\enm}{\end{bmatrix}}
\newcommand{\bs}[1]{{\boldsymbol #1}}
\newcommand{\T}{{\prime}}

\providecommand{\trace}{{\operatorname{tr}}}

\theoremstyle{definition}
\newtheorem{thm}{Theorem}

\newtheorem{defn}[thm]{Definition}

\newcommand{\E}{\mathbb{E}}

\newcommand{\Bm}{\begin{bmatrix}}
\newcommand{\Em}{\end{bmatrix}}

\newcommand{\forw}[1]{\mathcal{F}_{ #1}}

\newcommand{\backw}[1]{\mathcal{B}_{#1}}

\newcommand{\ho}[1]{{\color{DarkOrange} #1}}
\newcommand{\hb}[1]{{\color{RoyalBlue} #1}}

\newcommand*\circled[1]{%
  \tikz[baseline=(C.base)]\node[draw,circle,inner sep=0.5pt](C) {#1};\!
}

\tikzset{vert/.style={draw,circle, minimum size=6mm, inner sep=0pt, fill=white}}

\begin{document}

\maketitle

\begin{abstract}
\noindent
In this document I aim to give an informal treatment of automatic \ho{Backward Filtering Forward Guiding}, a general algorithm for conditional sampling from a Markov process on a directed acyclic graph. I'll show that the underlying ideas can be understood with a basic background in probability and statistics. The more technical treatment is the paper \cite{vandermeulen2021automatic}, which I will abbreviate to \hb{ABFFG}.  I specifically assume some background knowledge on  likelihood based inference and  Bayesian statistics. Section \ref{sec:cont} is more demanding: it assumes your are familiar with continuous-time stochastic processes constructed from their infinitesimal generator (see for instance the books by Liggett \cite{Liggett2010} or Bass \cite{bass2011stochastic}).  

\bigskip 

\emph{Clearly, all work discussed here is the result of research carried out over the past decade together with various collaborators, most importantly {\bf Moritz Schauer} (Chalmers University of Technology and University of Gothenburg, Sweden). Section \ref{sec:sde_tree} is based on joint work with  {\bf Marcin Mider} (Trium Analysis Online GmbH, Germany) and {\bf Frank Sch\"afer} (University of Basel, Switzerland) as well. }
\end{abstract}


Markov processes, and in particular state-space models, are among the most popular probabilistic constructions to model uncertainty in time-evolving data. The statistical problem consists of  extracting information about the process using observations from it.   For \hb{simple} settings \sidenote{Most notably linear Gaussian systems, where Kalman filtering has been central for over half a century}, it is known how to ``solve'' the statistical problem and the associated methods haven been implemented in mainstream engineering packages such as Matlab. I will start off  in Section \ref{sec:ssm} from the setting of state-space models, as I believe there is some chance of familiarity, which will ease digesting later generalisations. As we will see in Section \ref{sec:tree}, once the state-space case is well understood, some of these generalisations are almost straightforward.

\begin{figure}
\begin{center}
\begin{tikzpicture}

\tikzstyle{empty}=[fill=white, draw=black, shape=circle,inner sep=1pt, line width=0.7pt]
\tikzstyle{solid}=[fill=black, draw=black, shape=circle,inner sep=1pt,line width=0.7pt]

\begin{pgfonlayer}{nodelayer}
			\node [style=solid,label={$0$},] (0) at (-6, 0) {};
		\node [style=solid,label={$s$},] (1a) at (-4, -0.5) {};
		\node [style=solid,label={$t$},] (1b) at (-4.5, 2) {};
		\node [style=empty,label={$v_2$},] (2a) at (-3, 1) {};
		\node [style=solid,label={$u$},] (2b) at (-3, 3) {};
		\node [style=solid,label={$r$}] (2) at (-2, -0.5) {};
		\node [style=empty,label={$v_3$}] (3) at (-0, -0.5) {};

		\node [style=empty,label={$v_1$},] (4b) at (-1, 3) {};

\end{pgfonlayer}

\begin{pgfonlayer}{edgelayer}
		\draw [style=edge] (0) to (1a);
		\draw [style=edge] (0) to (1b);
		\draw [style=edge] (1b) to (2a);
		\draw [style=edge] (1b) to (2b);
		\draw [style=edge] (1a) to (3);

		\draw [style=edge] (2) to (3);
		\draw [style=edge] (2b) to (4b);
\end{pgfonlayer}

\end{tikzpicture}
\end{center} \caption{Example of a directed tree. The observations are denoted by $v_1$, $v_2$ and $v_3$. The values at all other vertices are unknown. \label{fig:tree_dag}}
\end{figure}
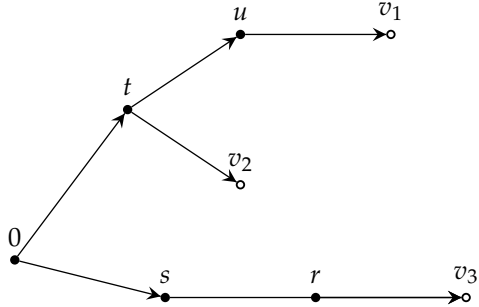

\newthought{However, before going there}, I'll discuss a visualisation of the general case I aim to deal with.
Consider a stochastic process on the tree in Figure \ref{fig:tree_dag}. 
Here, the root-vertex is depicted by $0$. Along each edge the process evolves according to either \hb{one step of a discrete-time Markov chain} or a \hb{time-span of a continuous-time Markov process}. 	At vertices $0$ and $t$ the process splits independently conditional on the values at $0$ and $t$ respectively. Observations are at the leaf-vertices $v_1$, $v_2$ and $v_3$. This setting encompasses  state-space models (popular for example in signal-processing and data-assimilation) and phylogenetic tree models arising in evolutionary biology. The statistical problem I consider consists of inferring the values of the process at the non-leaf vertices (i.e.\ $0$, $s$, $r$, $t$, $u$). Moreover, if the forward evolution over the edges depends on a parameter $\theta$, we may be interested in estimating $\theta$ as well.

\section{Likelihood computation for a state-space model}\label{sec:ssm}

Recall  a Markov process is a (time-evolving) memoryless process. 
This means that given the present state, the past state is irrelevant for its forward evolution.   A \hb{state-space model}\sidenote{Depending on the application area, state-space models and also referred to as \hb{hidden Markov models}.}  can  be depicted by the following diagram
\begin{center}
\begin{tikzpicture}[style={scale=0.8}]
	\tikzstyle{empty}=[fill=white, draw=black, shape=circle,inner sep=1pt, line width=0.7pt]
	\tikzstyle{solid}=[fill=black, draw=black, shape=circle,inner sep=1pt,line width=0.7pt]
	\begin{pgfonlayer}{nodelayer}
		\node [style=solid,label=below:{$0$},] (00) at (-6, 0) {};
		\node [style=empty,label={$v_0$},] (0obs) at (-6, 1.5) {};

		\node [style=solid,label=below:{$1$},] (0) at (-4, 0) {};
		\node [style=empty,label={$v_1$},] (1obs) at (-4, 1.5) {};

		\node [style=solid,label=below:{$2$}] (1) at (-2, 0) {};
		\node [style=empty,label={$v_2$},] (2obs) at (-2, 1.5) {};

		\node [] (2) at (-0, 0) {};

		\node [style=none] (end) at (1.0, 0) {.};

		\node [style=solid,label=below:{$n-1$}] (nn) at (2, 0) {};
		\node [style=empty,label={$v_{n-1}$},] (nnobs) at (2, 1.5) {};

		\node [style=solid,label=below:{$n$}] (n) at (4, 0) {};
		\node [style=empty,label={$v_{n}$},] (nobs) at (4, 1.5) {};

		\node [style=none] (end) at (1.0, 0) {};

	\end{pgfonlayer}
	\begin{pgfonlayer}{edgelayer}
		\draw [style=edge] (00) to (0);
		\draw [style=edge] (0) to (1);
		\draw [style=edge] (1) to (2);
		\draw [style=edge] (nn) to (n);

		\draw [style=edge] (00) to (0obs);
		\draw [style=edge] (0) to (1obs);
		\draw [style=edge] (1) to (2obs);
		\draw [style=edge] (nn) to (nnobs);
		\draw [style=edge] (n) to (nobs);

		\draw [style=dashed box] (2) to (nn);
	\end{pgfonlayer}
\end{tikzpicture}
\end{center}
Here both black dotted dots  and big open dots represent vertices of a graph. At each vertex resides a random quantity, which is either observed (open dot), or latent/non-observed (black filled dot).  
The arrow describes the  probabilistic evolution over an edge connecting two vertices. The arrows connecting the black dots constitute a graphical model for a latent (unobserved) Markov process.  If $x_s$ denotes the random quantity at vertex $s$, then the probability density of ``moving'' from $x_s$ to $x_t$ is denoted by $p(x_t \mid x_s)$, this is an instance of Bayesian notation \sidenote[][-1.0in]{If $f_Y$ denotes the density of the random quantity $Y$, then in fact we are talking about the mapping $y\mapsto f_Y(y)$. Bayesian notation means we simply write $p(y)$ here, omitting the subscript. This comes very handy at times, but one should be careful about $p(y^2)$, which is to be interpreted as $f_{Y^2}(y^2)$.  Later on I will denote  the Markov kernel connecting vertices $s$ and $t$ by $\kappa_{s\to t}$, rather than $p(x_t \mid x_s)$.}. As we number the black vertices by $0, 1, \ldots, n$, we have
\[ p(x_0, \ldots, x_n) = p(x_0) \prod_{i=1}^n p(x_i\mid x_{i-1}),  \]
which follows from the Markov property. 
Each observation, denoted by $v_i$, depends only on $x_i$ and it follows from the graphical structure that 
\[ p(v_0,\ldots, v_n \mid x_0,  \ldots, x_n) = \prod_{i=0}^n p(v_i \mid x_i). \]
Combining the previous two displayed formulas gives
\begin{fullwidth}
\begin{equation}\label{eq:lik}
\begin{split} p(v_0, \ldots, v_n ) & = \int p(v_0, \ldots, v_n \mid x_0, \ldots, x_n) p(x_0, \ldots x_n) \dd x_0 \cdots \dd x_n  \\  &= \int \left(p(x_0)\prod_{i=1}^n p(x_i\mid x_{i-1})\right) \left(
\prod_{i=0}^n p(v_i \mid x_i)\right) \dd x_0\cdots \dd x_n. 
\end{split}
\end{equation}
\end{fullwidth}
If the densities ``$p$'' appearing here depend on an unknown parameter $\theta$, then we can just add this as a subscript everywhere and we obtain a first result: the \ho{likelihood} for $\theta$ based on the observations $\mathcal{V}_n:= \{v_0,\ldots, v_n\}$ equals \sidenote{This is really a definition, the likelihood {\it is}  simply defined as the joint density of all observations. It is given a special name when viewed as a function of the parameter $\th$ for fixed observations, rather than the other way around. Note that as a function of $\theta$, the likelihood is just a nonnegative function: it is {\it not} a  density; it even need not be integrable. }
\[ L(\theta; \mathcal{V}_n) =   p_\theta(v_0, v_1, \ldots, v_n). \] 
Likelihood based inference then appears straightforward from here; depending on your preference for either  maximum likelihood or Bayesian inference, ``all'' is there.\sidenote[][1.5in]{Maximum likelihood means you determine $\argmax_{\theta \in \Theta} L(\theta; \mathcal{V}_n)$, $\Theta$ denoting the parameter set, and one can numerically carry out the optimisation. Bayesian inference additionally requires specification of a prior on $\theta$ and subsequently the likelihood and prior can be fed into a probabilistic programming language to produce samples from the posterior. Well known examples include \texttt{STAN} and \texttt{Turing}.} Hence, if all we care about is inferring $\theta$ and \eqref{eq:lik} can easily be evaluated, we're good. The issue is of course that \eqref{eq:lik} requires evaluation of an $n$-fold integral, which makes it kind of a beast.

One example when the likelihood can be evaluated in closed form is the linear Gaussian state-space model where\sidenote[][0.5in]{For this model, the Kalman filter provides the basis for numerically efficient evaluation of the likelihood.}
\begin{equation}\label{eq:linear_ssm} \begin{split}
 v_i \mid x_i & \sim N(L_i v_i, \Sigma_i) \\
 x_i \mid x_{i-1} & \sim N(B x_{i-1} + \beta, \Gamma_i)	
 \end{split}
\end{equation}
This tractability is lost if the second equation would for example read as
\begin{equation}\label{eq:nonlinear_ssm}	 x_i \mid x_{i-1}  \sim N(b(x_{i-1}), \Gamma_i), \end{equation}
with $x\mapsto b(x)$ a nonlinear map. If any of the distributions in \eqref{eq:linear_ssm} would be non Gaussian, then the calculation would also break down.

\newthought{Besides parameter estimation we may also be interested} in recovering the latent states $\bs{x}:=(x_0,\ldots, x_n)$. For example, when $v_0, v_1,\ldots, v_n$ is a noisy version of an underlying signal $x_0, x_1,\ldots, x_n$, or when the observations $v_i$ only measure part of the signal $x_i$.  I will take the Bayesian point of view here, which means that I view $\prod_{i=0}^n p(v_i \mid x_i)$ as the likelihood (it comes from the observation equation in the state-space model) and $p(x_0)\prod_{i=1}^n p(x_i\mid x_{i-1})$ as the prior density of $\bs{x}$ (it comes from the state equation in the state-space model). Then, we wish to find the posterior density
\begin{fullwidth}
\[ p^\star(x_0,\ldots, x_n):= p(x_0, \ldots, x_n\mid v_0,\ldots, v_n) = \frac{p(x_0)\prod_{i=1}^n p(x_i\mid x_{i-1}) \times \prod_{i=0}^n p(v_i \mid x_i)}{p(x_0)\int \prod_{i=1}^n p(x_i\mid x_{i-1}) \times \prod_{i=0}^n p(v_i \mid x_i) \dd x_i}. \] 
\end{fullwidth}
Note that $p(x_0)$ cancels out. 
The great thing about Markov Chain Monte Carlo methods is that we don't need to evaluate the denominator. So if there is no unknown parameter, this looks good.

\newthought{Finally, think about the actual setting} we often encounter in practice: the parameter is unknown and we wish to infer \ho{both $\bs{x}$ and $\theta$}. What to do? And yes, keep in mind that later I wish to extend to the setting where the arrows on the edges correspond to evolving a continuous-time Markov
process for some time interval.
\begin{marginfigure}
\includegraphics[scale=0.5]{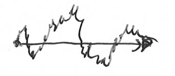}
	\caption{Transition over an edge according to a continuous-time Markov process with ``diffusion''-behaviour.}
\end{marginfigure}

\section{Backward Information Filter (BIF)}
As said, evaluation of \eqref{eq:lik} is not trivial (in fact, impossible for most models). 
 To deal with this problem, the first thing to notice is that there is an efficient recursive way to compute it. This may remind you of dynamic programming, what I explain here is a simple version of the product-sum algorithm which is well explained in Chapter 8 of \cite{Bishop07} for example.
The idea is to compute the \ho{$h$-function} \sidenote{The terminology $h$-function is nonstandard. We borrow it from the much related concept of Doob's $h$-transform.}
\begin{equation}\label{eq:h} h(x_i) = p(v_i,\ldots, v_n \mid x_i).\end{equation} For $i=n$ this is simple: $h(x_n) = \ho{p(v_n \mid x_n)}$.
\begin{fullwidth}
Now note the following recursive relation
	\begin{align*}
	p(v_n, v_{n-1} \mid x_{n-1}) & = \int p(v_n, v_{n-1}, x_n \mid x_{n-1}) \dd x_n \\ &= \int p(v_n, v_{n-1} \mid x_n, x_{n-1}) p(x_n \mid x_{n-1}) \dd x_n \\ &	= p(v_{n-1} \mid x_{n-1}) \int \ho{p(v_n \mid x_n)} p(x_n \mid x_{n-1}) \dd x_{n}.
	\end{align*}
\end{fullwidth}
Denoting the left-hand-side by $h(x_{n-1})$ this reads
\begin{equation}\label{eq:rec_h} h(x_{n-1}) =  p(v_{n-1} \mid x_{n-1}) \int h(x_n) p(x_n \mid x_{n-1}) \dd x_{n}. \end{equation}
The ``n'' can in fact be replaced by $i$ and this recursion is known as the \hb{Backward Information Filter} (BIF).\sidenote[][-0.5in]{The BIF can be applied more generally on a directed tree and, with some adaptation, also on a Directed Acyclic Graph (DAG).}
\begin{fullwidth}
\begin{equation*}
	\underbrace{p(v_n \mid x_n)}_{\displaystyle h(x_n)}\quad  \longrightarrow\quad  \underbrace{p(v_n, v_{n-1} \mid x_{n-1})}_{\displaystyle h(x_{n-1})}\quad  \longrightarrow \quad \cdots\quad \longrightarrow\quad  \underbrace{p(v_n,\ldots, v_0 \mid x_0)}_{\displaystyle h(x_0)}
\end{equation*}	
\end{fullwidth}
The notation I use here is rather informal \sidenote{Here, Bayesian notation starts to break-down, also as I apply it to $h$, so $h(x_n)$ is in fact $h_n(x_n)$ and similarly $h(x_{n-1})$ is $h_{n-1}(x_{n-1})$.}. Equation \eqref{eq:rec_h} can be viewed as follows: at time $n-1$ there are two children vertices: the observation at time $n-1$ and the vertex corresponding to $x_n$. The leaf vertex gives as contribution $p(v_{n-1} \mid x_{n-1})$ while the vertex for $x_n$ gives contribution  $\int h(x_n) p(x_n \mid x_{n-1}) \dd x_{n}$. Further ahead we will call the latter the \ho{pullback} of $h$ along $p(x_n \mid x_{n-1})$. Finally, both child contributions are multiplied to arrive at \eqref{eq:rec_h}. 

The terminology ``Backward Information Filter'' is perhaps only partially appropriate.  It is an algorithm with steps running backwards in time taking the data (``information'') into account, so calling it ``Backward Information'' seems appropriate.  ``Filter'' may be a bit confusing, because commonly the filtering density of state of $x_i$ (say) is defined by $p(x_i \mid v_0,\ldots v_i)$. The BIF is about computing $p(v_i,\ldots, v_n \mid x_i)$ though. 

Now suppose $h_i$ has been computed (suppose we can actually do this for now). Define
\begin{equation}\label{eq:star} p^\star(x_i \mid x_{i-1}) = \frac{p(x_i \mid x_{i-1})h(x_i)}{\int p(x_i \mid x_{i-1})h(x_i) \dd x_i}. \end{equation}
What is this density reflecting? Assume at time $i-1$ you know $x_{i-1}$ but can also peak into the future and see $v_i,\ldots, v_n$ (this is the case: these are part of the observed data). Then $p^\star(x_i \mid x_{i-1})$ is the density of moving to $x_i$ in view of this information. \marginnote{Note that $p^\star$ is obtained by a change of measure on $p$ using $h$. This transform is known as \ho{Doob's $h$-transform}.} The $^\star$ reminds us of conditioning on $v_i,\ldots, v_n$. Plugging the parameter $\theta$ back into the notation, and assuming prior distribution $p(\theta)$ for the parameter, we can sample from $\theta, \bs{x} \mid \scr{V}_n$ by the following iterative scheme\sidenote{This is the Gibbs sampler, in this setting also known as data-augmentation. The algorithm requires initialisation of $\theta$ or, if the first step consists of sampling $\theta$,  $\bs{x}$.}
\begin{itemize}
	\item sample $\bs{x} \mid \theta, \scr{V}_n$; the ``target'' density being proportional to $\prod_{i=0}^n p^\star_\th(x_i \mid x_{i-1})$;
	\item sample $\theta \mid \bs{x}, \scr{V}_n$; the ``target'' density being proportional to $p(\theta)\prod_{i=0}^n p^\star_\th(x_i \mid x_{i-1})$.
\end{itemize}
Here, for $i=0$,  $p(x_0 \mid x_{-1})$ is simply meant to be $p(x_0)$, simplifying notation.

\newthought{What did we obtain so far?} We recursively compute $h$ as in \eqref{eq:h} and derived a two-step sampling procedure to sample from the \ho{joint} distribution of hidden states $\bs{x}$ and parameter $\theta$. All of this works, provided we can actually compute $h$.


\section{Forward guiding}
There are few cases where $h$ can be computed in closed form, the easy cases include
\begin{enumerate}
	\item the \hb{discrete} setting, where $x_i,\, v_i \in E$ and $E$ can be represented by the set of labels $E=\{1,\ldots, R\}$;
	\item the \hb{linear Gaussian setting}, where $x_i \mid x_{i-1} \sim N(B_i x_{i-1} + \beta_i, \Gamma_i)$ and $v_i \mid x_i \sim N(L_ix_i, \Sigma_i)$. 
\end{enumerate}
Now imagine $p(x_i \mid x_{i-1}) \approx \tilde p(x_i \mid x_{i-1})$ and $p(v_i \mid x_i) \approx \tilde{p}(v_i\mid x_i)$, where $\tilde p$ falls in one of the two enumerated settings. An initial thought could be: ``Ok, let's use the approximation $\tilde p$ then, with a bit of luck this is not too bad.''. In fact, we can (and should) do better. \sidenote{This is an important point which I have often seen misunderstood. As an example, consider the state-space model where the state evolves according to \eqref{eq:nonlinear_ssm}. By linearisation, we may be able to find $(B,\beta)$ in \eqref{eq:linear_ssm} which would then define $\tilde p$.} What we rather propose to do, is performing the BIF with $\tilde p$, yielding maps $g$ (this is tractable, by \ho{choice} of $\tilde p$) and defining 
\begin{equation}\label{eq:circ} p^\circ(x_i \mid x_{i-1}) = \frac{p(x_i \mid x_{i-1}) g(x_i)}{\int p(x_i \mid x_{i-1}) g(x_i) \dd x_i}. \end{equation}
Note that this resembles the definition of $p^\star$ in \eqref{eq:star}. Whereas $h$ in \eqref{eq:star} ensures correct conditioning, $g$ in \eqref{eq:circ} ensures \hb{guiding} to take the observations into account. \sidenote{Put differently, $p^\circ$ is obtained using  Doob-$h$-transform with $g$, just like $p^\star$ is obtained with $h$.} Note that $p$ is still in the expression for $p^\circ$! The process that evolves under \eqref{eq:circ} is called the \hb{guided process}. 
  
 \newthought{The reason that this is useful} lies in the fact that we can compute the likelihood ratio between the star and circ densities.
 Clearly, 
 \begin{equation}\label{eq:pstar/pcirc} \frac{p^\star(x_1,\ldots, x_n)}{p^\circ(x_1,\ldots, x_n)} = \prod_{i=1}^n \frac{h(x_i)}{g(x_i)} \frac{\int p(x_i \mid x_{i-1}) g(x_i) \dd x_i}{\int p(x_i \mid x_{i-1}) h(x_i) \dd x_i}. \end{equation}
 Using the recursive relation \eqref{eq:rec_h} this can be simplified. Without loss of generality, assume $x_0$ to be known and drop the observation $v_0$.\sidenote{We can always add an artificial root-node and then edges originating from this root node represent a prior distribution on initial states.} As $h$ and $g$ satisfy the BIF for $p$ and $\tilde p$ respectively, we have for $2\le i \le n$
 \begin{align*}
 	h(x_{i-1}) &= p(v_{i-1} \mid x_{i-1}) \int h(x_i) p(x_i \mid x_{i-1} \dd x_i \\
 	 	g(x_{i-1}) &= \tilde p(v_{i-1} \mid x_{i-1}) \int g(x_i) \tilde p(x_i \mid x_{i-1} \dd x_i 
 \end{align*}
 Substituting these expressions into \eqref{eq:pstar/pcirc} gives
 \begin{fullwidth}
 \begin{align*}
 	 \frac{p^\star(x_1,\ldots, x_n)}{p^\circ(x_1,\ldots, x_n)} &= \frac{h(x_n)}{g(x_n)} \left(\prod_{i=2}^n \frac{p(v_{i-1} \mid x_{i-1})}{\tilde p(v_{i-1} \mid x_{i-1})}  \frac{\cancel{\int p(x_i \mid x_{i-1})  h(x_i) \dd x_i}}{\int \tilde p(x_i \mid x_{i-1}) g(x_i) \dd x_i} \frac{\int p(x_i \mid x_{i-1}) g(x_i) \dd x_i}{\cancel{\int p(x_i \mid x_{i-1}) h(x_i) \dd x_i}}\right)  \\ & \qquad  \times \frac{\int p(x_1 \mid x_0) g(x_1) \dd x_1}{\int p(x_1 \mid x_0)  h(x_1) \dd x_1}  \\ & =\left(\prod_{i=1}^n \frac{p(v_i\mid x_i)}{\tilde p(v_i\mid x_i)}\right) \left(\prod_{i=2}^n \frac{\int  p(x_i \mid x_{i-1}) g(x_i) \dd x_i}{\int \tilde p(x_i \mid x_{i-1}) g(x_i) \dd x_i} \right)\frac{g(x_0)}{h(x_0)}.
 \end{align*}	
 \end{fullwidth}
If we let 
\[ \Psi(x_1,\ldots, x_n) :=  \left(\prod_{i=2}^n \frac{\int  p(x_i \mid x_{i-1}) g(x_i) \dd x_i}{\int \tilde p(x_i \mid x_{i-1}) g(x_i) \dd x_i} \right)\left(\prod_{i=1}^n \frac{p(v_i\mid x_i)}{\tilde p(v_i\mid x_i)}\right), \]
then this can be rewritten to 
 \begin{equation}\label{eq:lr}  \frac{p^\star(x_1,\ldots, x_n)}{p^\circ(x_1,\ldots, x_n)} =\frac{g(x_0)}{h(x_0)}
  \Psi(x_1,\ldots, x_n). \end{equation}
Whereas $p^\star$ is intractable (because $h$ is), we have $p^\circ$ at our disposal and impose the assumption that sampling from  $\bs{x}$ under $p^\circ$ is tractable. The above formula tells us how to correct for the discrepancy between $p^\star$ and $p^\circ$. 
 In fact, \hb{everywhere we encounter $p^\star(x_1,\ldots, x_n)$ we can safely replace it with $p^\circ(x_1,\ldots, x_n)$ times the product on the right-hand-side of \eqref{eq:lr}}. The beauty of the shown derivation lies in the observation that  the $h(x_i)$ almost all cancel.  The only intractable term $h(x_0)$ fortunately turns out to cancel in Markov Chain Monte Carlo methods!
 
 Alternatively, we can multiply both sides of Equation \eqref{eq:lr} by $p^\circ(x_1,\ldots, x_n)$ and then integrate over $(x_1,\ldots, x_n)$. This implies
 \[ h(x_0)=g(x_0)\, \EE^\circ \Psi(X_1,\ldots, X_n). \]
 The left-hand-side is the likelihood and the expression shows how it can be obtained from $g(x_0)$ multiplied by an expectation of a path-functional of the guided process. 
 
 
 \section{Backward Filtering Forward Guiding}
 
 This section can be a short: we just combine what we have derived. That is, we use $\tilde p$ for the BIF to get $g$. This defines $p^\circ$  via \eqref{eq:circ}. Then we can forward sample $\bs{x}$ under $p^\circ$ to \hb{guide}  $\bs{x}$ to the observations and  compute a correction by \eqref{eq:lr}. So what we do is \ho{backward filtering}, followed by \ho{forward guiding}. 
 \sidenote[][-0.5in]{A natural question is whether one could also do forward filtering, backward guiding. While in certain cases this is indeed possible,  forward guiding  is more practical, because is shares structure with the unconditional forward dynamcis.
  }
 
 Classical cases, where actually the forward model corresponds to the discrete or linear Gaussian setting, are special cases. In such settings we don't need to use an approximate $g$ (however, it still can be computationally advantegeous). If we don't use the approximation, then $p^\star=p^\circ$, and the right-hand-side of \eqref{eq:lr} will be $1$. Then, if we only care about parameter estimation, there is no need to do forward guiding: the BIF will result in a closed form expression for the likelihood which may subsequently be used in likelihood based inference. 
 However, as in a general setting it will be impossible to compute the BIF filter efficiently, performing the BIF for a simpler process  will be a way out. Let me stress again that the guided process still contains the (possibly complicated) forward transition density $p$. Note that due to the Markov property we only need to be able to sample one step forward according to $p^\circ$, the BIF-backward recursion is inherently more difficult. 
 
  \newthought{One way to view the combined procedure} of backward filtering forward guiding is as follows: we compute $g(x_n)$ and put it on a pile. Next, we compute $g(x_{n-1})$ and put it on top of that pile. We continue until we get $g(x_0)$. In the end, we have a pile with (from top to bottom) \begin{equation}\label{eq:pile_h} [g(x_0), g(x_1),\ldots, g(x_n)]. \end{equation} Next to it, we place the pile with (again from top to bottom)  \begin{equation}\label{eq:pile_p} [p(x_0), p(x_1\mid x_0),\ldots, p(x_n \mid x_{n-1}]. \end{equation} 
 Then we simply pick the top element from both piles, combine the contributions from each pile into $p^\circ$ and simulate from it to get $x_0^\circ$. Repeating this procedure until the pile is empty results in the samples
 \[ [x_0^\circ, x_1^\circ, \ldots, x_n^\circ]. \]
 
\section{Extension to a tree and general DAG}\label{sec:tree}

The state-space model considered so far has a very simple topology. In what follows, I'll generalise the approach to a tree topology. This means that at any vertex, there can be multiple leaf vertices, and that any vertex may ``duplicate'' followed by conditionally independent evolutions over both duplicates. 
To explain the setting, consider the typical setting depicted in  Figure \ref{fig:tree}.
\begin{marginfigure}
\begin{center}
\begin{tikzpicture}

\tikzstyle{empty}=[fill=white, draw=black, shape=circle,inner sep=1pt, line width=0.7pt]
\tikzstyle{solid}=[fill=black, draw=black, shape=circle,inner sep=1pt,line width=0.7pt]

\begin{pgfonlayer}{nodelayer}
		\node [style=solid,label={$s$},] (0) at (-4, 3.5) {};

		\node [style=solid,label={$t_1$},] (2a) at (-2, 4.5) {};
		\node [style=empty,label={$u$},] (2b) at (-2, 3.5) {};
		\node [style=solid,label={$t_2$}] (2) at (-2, 2.5) {};
\end{pgfonlayer}

\begin{pgfonlayer}{edgelayer}

		\draw [style=edge] (0) to (2);
		\draw [style=edge] (0) to (2a);
		\draw [style=edge] (0) to (2b);
\end{pgfonlayer}

\end{tikzpicture}
\end{center}
	\caption{Part of a tree with parent vertex $s$.}
	\label{fig:tree}
\end{marginfigure}
The vertex labeled $s$ has three children: $t_1$, $t_2$ and $u$. As before we assume the Markov property, meaning that $x_{t_1}$, $x_{t_2}$ and $u$ are independent, conditional on $x_s$. We then have
\[ h(x_s) = p(u \mid x_s) \prod_{i=1}^2 \int p(x_{t_i} \mid x_s) h(x_{t_i}) \dd x_{t_i}. \]
This can be viewed as each of the children, $t_1$, $t_2$ and $u$, sending a \hb{message} to their common parent vertex. After vertex $s$ has received messages from all of its children, the messages get multiplied. Indeed, BFFG can be interpreted as a message passing algorithm with messages (for this specific example)
\begin{align*}
 m_{t_i}(x_s) &= 	\int p(x_{t_i} \mid x_s) h(x_{t_i}) \dd x_{t_i},\qquad i=1, 2, \\
 m_{u}(x_s) & = p(u \mid x_s).
\end{align*}

\newthought{For a general Directed Acyclic Graph (DAG)} there is one additional ingredient needed. The difficulty lies in the fact that a vertex can have multiple parent vertices.\sidenote{As an example, suppose at a vertex we compute the sum of the values at the parent indices.} For such a vertex we need to ``split'' $h$  in the backward filtering step to its parents. Hence, as an example, we need to decompose $h(x_1, x_2)$  into $h_1(x_1)$ and $h_2(x_2)$. A tractable approach for doing this is in the \hb{ABFFG}-paper. This is a bit of an opposite operation compared to fusion, though whereas fusion is exact, an approximation is made when doing  a split operation in backward filtering. Nevertheless, as explained in the paper, we can devise an algorithm for sampling from the exact smoothing distribution.  


\section{A toy example for the Backward Information Filter on a tree}

\begin{figure}[h]
\begin{center}
\begin{tikzpicture}

\tikzstyle{empty}=[fill=white, draw=black, shape=circle,inner sep=1pt, line width=0.7pt]
\tikzstyle{solid}=[fill=black, draw=black, shape=circle,inner sep=1pt,line width=0.7pt]

\begin{pgfonlayer}{nodelayer}
		\node [style=solid,label={$x_{-1}$},] (root) at (-8, 0) {};
		\node [style=solid,label={$x_{0}$},] (0) at (-6, 0) {};
		\node [style=solid,label={$x_{1}$},] (1a) at (-4, -0.5) {};
		\node [style=solid,label={$x_3$},] (1b) at (-4.5, 2) {};
		\node [style=empty,label={$v_2$},] (2a) at (-3, 1) {};
		\node [style=solid,label={$x_4$},] (2b) at (-3, 3) {};
		\node [style=solid,label={$x_2$}] (2) at (-2, -0.5) {};
		\node [style=empty,label={$v_3$}] (3) at (-0, -0.5) {};
		\node [style=empty,label={$v_1$},] (4b) at (-1, 3) {};

		\node [style=morphism] (kap_-1_0) at (-7, 0) {\hb{$\kappa_{-1,0}$}};
		\node [style=morphism] (kap_0_t) at (-5.25, 1) {\hb{$\kappa_{0,3}$}};
		\node [style=morphism] (kap_0_s) at (-5.25, -0.25) {\hb{$\kappa_{0,1}$}};
		\node [style=morphism] (kap_s_r) at (-3, -0.5) {\hb{$\kappa_{1,2}$}};
		\node [style=morphism] (kap_r_v3) at (-1, -0.5) {\hb{$\lambda_3$}};
		\node [style=morphism] (kap_t_v2) at (-3.75, 1.5) {\hb{$\lambda_2$}};
		\node [style=morphism] (kap_t_u) at (-3.75, 2.5) {\hb{$\kappa_{3,4}$}};
				\node [style=morphism] (kap_r_v3) at (-2, 3) {\hb{$\lambda_1$}};
\end{pgfonlayer}

\begin{pgfonlayer}{edgelayer}
		\draw [style=edge] (root) to (0);
		\draw [style=edge] (0) to (1a);
		\draw [style=edge] (0) to (1b);
		\draw [style=edge] (1b) to (2a);
		\draw [style=edge] (1b) to (2b);
		\draw [style=edge] (1a) to (3);
		\draw [style=edge] (2) to (3);
		\draw [style=edge] (2b) to (4b);
\end{pgfonlayer}

\end{tikzpicture}
\end{center}
\caption{Copy of Figure \ref{fig:tree_dag}, though with an extra vertex $-1$, and the kernels along edges added.\label{fig:tree_dag_withkernels}}
\end{figure}
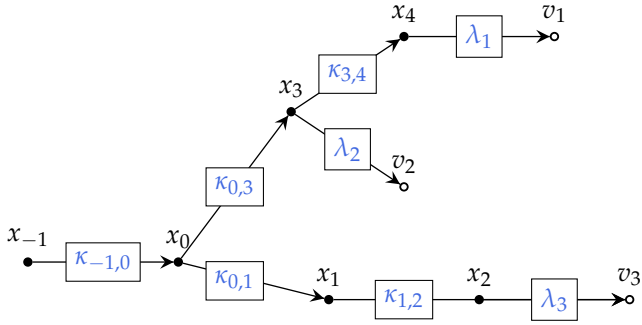

We illustrate the BIF for the directed acyclic graph depicted in Figure \ref{fig:tree_dag}.  
As we do not observe the value at vertex $0$, we equip this with a prior. This essentially means that we include an artificial vertex, which we denote by $-1$, pointing towards $0$. Furthermore, we make the kernels along the edges explicit to obtain Figure \ref{fig:tree_dag_withkernels}.

Suppose $x_t \in E:=\{\circled{1},\,\circled{2},\,\circled{3}\}$ and $v_t \in \{	\circled{1,2}\,, \circled{3}\}$.  The idea is that in observing we cannot  distinguish $\circled{1}$ and $\circled{2}$.
As the state-space is finite, we can identify  Markov kernels with transition kernels. To make the example a bit more explicit, suppose 
\[ \lambda_i =\begin{bmatrix}
	1 & 0 \\ 1& 0 \\ 0 & 1
\end{bmatrix} \qquad \kappa_{s,t}=\begin{bmatrix} 1-\theta & \theta & 0 \\
0.25 & 0.5 & 0.25  \\ 0.4 & 0.3 & 0.3 \end{bmatrix},\]
for $i \in \{1,2,3\}$, $s\in \{0,1,3\}$ and $t\in \mbox{ch}(s)$ (meaning vertex $t$ is a child of vertex $s$). There is an unknown prior $\th\in [0,1]$ in the matrix $\kappa_{s,t}$, which is the probability to go from state $\circled{1}\,$ to $\circled{2}\,$. 
The prior on the initial state is defined by setting  $x_{-1} = \circled{0}\:$\sidenote{It is completely irrelevant what the state of $x_{-1}$ is.} and \[
 \kappa_{-1,0} = [\pi_1,\: \pi_2,\: \pi_3]=:\bs{\pi}.\]	
 Thus, $\bs{\pi}$ contains the prior probabilities on $x_0$. 
To compute the BIF, first note that since the state space is finite, the map $x\mapsto h_t(x)$ can be identified with the (column)vector $h_t =[h_t(\circled{1}\,), h_t(\circled{2}\,), h_t(\circled{3}\,)]$.
We initialise from observations: for $t\in \{1, 2, 3\}$
\[ h^{\mathrm{obs}}_t := \begin{bmatrix} 1\\ 0\end{bmatrix} \ind\{v_t = \circled{1,2}\,\} +   \begin{bmatrix} 0\\ 1\end{bmatrix} \ind\{v_t = \circled{3}\,\}. \] 
Now let's start computing $h$ recursively towards the roots, starting from $v_3$. I claim
\[ h_2 = \lambda_3 h_3^{\mathrm{obs}} \qquad \qquad   h_1 = \kappa_{1,2} h_2. \]
In other words, computing $h_2$ and $h_1$ simply follows from taking matrix-vector products. 
To see why $h_1 = \kappa_{1,2} h_2$ is correct, note that for $x_1 \in E$ \sidenote{The $p(v_3 \mid x_1, x_2)=p(v_3 \mid x_2)$ by the Markov property.}
\begin{align*} h_1(x_1)=p(v_3 \mid x_1) =& \sum_{x_2 \in E} p(v_3, x_2 \mid x_1)  \\   =&    \sum_{x_2 \in E} \underbrace{p(v_3 \mid \cancel{x_1}, x_2)}_{\displaystyle h_2(x_2)} p(x_2\mid x_1) . \end{align*}
Now at vertices $0$ and $3$ there is a split, and we need to think how to deal with this case. Let's focus on vertex $3$:
\begin{center}
\begin{tikzpicture}

\tikzstyle{empty}=[fill=white, draw=black, shape=circle,inner sep=1pt, line width=0.7pt]
\tikzstyle{solid}=[fill=black, draw=black, shape=circle,inner sep=1pt,line width=0.7pt]

\begin{pgfonlayer}{nodelayer}
		\node [style=solid,label={$x_{-1}$},] (root) at (-8, 0) {};
		\node [style=solid,label={$x_{0}$},] (0) at (-6, 0) {};
		\node [style=solid,label={$x_{1}$},] (1a) at (-4, -0.5) {};
		\node [style=solid,label={$x_3$},] (1b) at (-4, 1) {};
		
		\node [style=morphism] (kap_-1_0) at (-7, 0) {\hb{$\kappa_{-1,0}$}};
		\node [style=morphism] (kap_0_s) at (-5.25, -0.25) {\hb{$\kappa_{0,1}$}};
		\node [style=morphism] (kap_0_t) at (-5.25, 0.5) {\hb{$\kappa_{0,3}$}};
\end{pgfonlayer}

\begin{pgfonlayer}{edgelayer}
		\draw [style=edge] (root) to (0);
		\draw [style=edge] (0) to (1a);
		\draw [style=edge] (0) to (1b);

\end{pgfonlayer}

\end{tikzpicture}
\end{center}
Just as before, we get\sidenote{We have to be a bit more careful in the notation here, for otherwise we would have obtained two different definition of $h_0$. For this reason, $h$ sent to vertex $0$ originating from vertex $3$ is denoted by $h_{0\pf 3}$.}
\[	 h_{0\pf 3} = \kappa_{0,3} h_3 \quad \text{and} \quad  h_{0\pf 1} = \kappa_{0,1} h_1.\]
As the forward path evolves conditionally independent, given the value of $x_3$, we have 
\[ h_0(x) = h_{0\pf 1}(x) h_{0\pf 3}(x). \]
This combination of $h_{0\pf 1}$ and  $h_{0\pf 3}$ we call \ho{fusion}.  
As the maps $x\mapsto h_t(x)$ can be identified with vectors, this implies that 
\[ h_0 =  h_{0\pf 1} \odot  h_{0\pf 3}, \]
with $\odot$ denoting the Hadamard (entrywise) product. 
In this way we could even write down the likelihood in terms of matrix-vector products and entrywise vector products: \sidenote{We have hidden dependence of the matrices $\kappa$ on $\theta$ to alleviate notation, but except for $h_i^{\mathrm{obs}}$, all $h$-vectors depend on $\theta$.}
\[
\begin{split}
 h_{0\pf 3} & = \kappa_{0,3}\left( (\kappa_{3,4}\lambda_1 h_1^{\mathrm{obs}}) \odot (\lambda_2 h_2^{\mathrm{obs}})\right) \\
h_{0\pf 1} &= \kappa_{0,1}\kappa_{1,2} \lambda_3 h_3^{\mathrm{obs}}\\
L(\th) & = \kappa_{-1,0} \left(h_{0\pf 1} \odot h_{0\pf 3}\right)
	\end{split}
\]
Note however that the separate steps, where we traverse the tree in backwards order, are much more insightful. 

\newthought{From this example we learn } that the BIF consists of composing the calculations  $\kappa h$ and $h_1 \odot h_2$. The first of these, we will call \ho{pullback} of $h$ along $\kappa$ (to be defined in more generality in the upcoming section), while the latter we called \ho{fusion}.

\section{Compositionality}

Reading code not written yourself is often hard. Even pseudo-code, as sometimes seen in scientific papers I find usually hard to digest. Especially in filtering, there appear so many indices! 
Older versions of the \hb{ABFFG} manuscript also contained those indices, but in fact we can get rid of those. Key is \hb{compositionality}: assembling the bigger, more complex algorithm by piecing together smaller, simpler pieces. 
That is exactly what we can do here: first we formalise our notation a bit. We assume that each forward transition corresponds to a Markov kernel $\kappa(x,\dd y)$ \sidenote{This means that for a (measurable) set $B$, the mapping $x\mapsto \kappa(x,B)$ is measurable and that for fixed $x$, $B\mapsto \kappa(x,B)$ ia a probability measure. The idea is that if at time $i$ the process is at $x$, then the state at time $i+1$ is drawn from the measure $\kappa(x,\cdot)$. It the state-space is finite, this simply boils down to sampling the state from a (finite) probability vector.}

For a Markov kernel we have the following two linear operators. For a bounded measurable function $h$ we define the \ho{pullback} by \begin{equation}\label{eq:pullback}(\kappa h)(x) = \int_E \kappa(x,\dd y) h(y).\end{equation}
To give this a probabilistic interpretation, note that $(\kappa h)(x) = \EE [ h(X_{n+1}) \mid X_n =x]$. As an example, if the state space is finite (say $E=\{1,\ldots, R\}$), then the preceding display reads  $(\kappa h)(x) = \sum_{y=1}^R \kappa(x,y) h(y)$ and $\kappa(x,y)$ is the one-step transition probability to go from state $x$ to $y$. We actually used this in the example of the previous section.

 For a measure $\mu$ we define \[ (\mu \kappa)(\dd y) = \int \mu(\dd x)\kappa(x,\dd y).\]
 This is the \ho{pushforward} of the measure $\mu$. The interpretation is as follows: suppose at time $n$ we sample $x_n$ from the measure $\mu$ and subsequently evolve the Markov chain from $x_n$ to $x_{n+1}$ according to the Markov kernel $\kappa$. Then $\mu\kappa$ is the distribution of $x_{n+1}$\sidenote{We first compute the joint distribution of $(x_n, x_{n+1})$ and then integrate out $x_n$.}. In the finite-state setting we have that for $x\in E$, $(\mu \kappa)(x) = \sum_{x=1}^R \mu(x) \kappa(x,y)$. 
 
 Recall in the description of BFFG the analogy of having the two piles \eqref{eq:pile_h} and \eqref{eq:pile_p}. This analogy can be formalised as viewing one  step of BFFG  as applying a \ho{backward} map together with a \ho{forward} map.
 
Recall that in each step of the BIF  we take a function $h(x_i)$ and do two things:
\begin{itemize}
	\item we put it on top of the ``$h$-pile'' \eqref{eq:pile_h};
	\item we compute $h(x_{i-1})$ as in \eqref{eq:rec_h} (note that part of this computation is indeed the pullback as defined in \eqref{eq:pullback}).
\end{itemize}
We interpret dropping $h(x_i)$ on the $h$-pile as \hb{sending a message $m$} which is used later in forward sampling (guiding). 
Viewed a bit more abstractly, each step in the BIF takes a function $h$, produces a new function  $h'$ and sends a message $m$.  Once all backward steps of the BIF have been completed, we have the pile of messages and we can combine it with the pile of forward evolutions, alike \eqref{eq:pile_p}. More formally we will shortly define  a forward map for this. 
 
Before entering the definitions, let's look at a small visualisation:
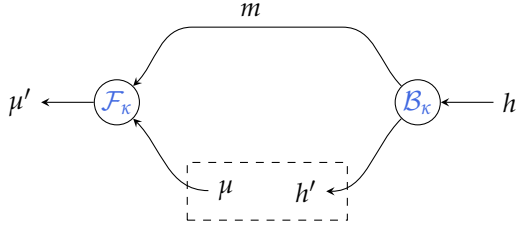
\begin{figure}
\begin{center}
\begin{center}
\begin{tikzpicture}
    \node[vert] (forw) at (0, 0) {\hb{$\mathcal{F}_\kappa$}};
    \node[vert] (backw) at (4, 0) {\hb{$\mathcal{B}_\kappa$}};

    \node (muout) [left =0.7 of  forw] {$\mu'$};
    \node (mu) [below right = 0.7 and 1 of forw] {$\mu$};
    \node (h) [right =0.7 of  backw] {$h$};
    \node (backwh) [below left = 0.7 and 1 of backw] {$h'$};
	\node (m) [above right =0.8 and 1.3 of forw] {$m$};

    \draw[<-] (muout) -- (forw);
    \draw[<-] (forw) to[out=south east, in=west] (mu) ;

    \draw[->] (h) -- (backw);
    \draw[->] (backw) to[out=south west,in=east] (backwh);
     \node[draw,dashed,fit=(mu) (backwh), inner xsep = 8pt] (box) {};
     
  \draw[<-] (forw) to[out=north east, in=west] ++(1,1)
     to ++(2,0)
     to[out=east, in=north west] (backw)
    ;


\end{tikzpicture}\\[0.2in]

\end{center}

	\caption{One step of BFFG. Read from right to left. For Definition 3, $\mathcal{B}_{\kappa}$ needs to be replaced by $\mathcal{B}_{\tilde\kappa}$.}
\end{center}
\end{figure}
We start from the right, where $h$ serves as input to a backward map $\mathcal{B}$. This map produces $h':= \kappa h$, but also a \ho{message $m$}, which is used in the forward map $\mathcal{F}$. The latter pushes forward the measure $\mu$ using the message $m$.

\begin{defn}\label{def:backwardmap} For a Markov kernel $\kappa$ and function $h$ define the \hb{backward map} $\backw\kappa$ by \sidenote{Compatibility of $\kappa$ and $h$ is implicitly assumed.}
\begin{equation}\label{eq:backw}
\backw{\kappa}(h) = \left(m, \kappa h\right),\quad \text{where} \quad  m(x,y) = \frac{h(y) }{(\kappa h)(x)}.
\end{equation}
\end{defn}
This map returns both the  pullback $\kappa h$  and an appropriate \hb{message} $m$ for the map $\forw\kappa$ specified in the following definition.
\begin{defn}\label{def:forwardmap}
For a Markov kernel $\kappa$, message $m$ (as defined in \eqref{eq:backw}) and measure $\mu$ define the \hb{forward map} $\forw\kappa$ by \sidenote{Again, compatibility of  $\kappa$, $m$ and $\mu$ is implicitly assumed. }
\begin{equation}\label{eq:forw}
 \forw{\kappa}(m, \mu)  = \nu, \quad \nu(\!\dd y) = \int m(x,y) \mu(\!\dd x) \kappa(x, \dd y). 
 \end{equation}
\end{defn}


\newthought{If $\mu$ is a probability measure} and $\backw\kappa$ sends the message $m$, then $\forw\kappa(m,\mu)$ is again a probability measure. 
If the BIF is intractable, we replace $\kappa$ in the backward map by the kernel $\tilde\kappa$, where $\tilde\kappa$ is chosen such that the BIF \textit{is} tractable.\sidenote{The indices $\kappa$ and $\tilde\kappa$ reflect the true forward dynamics and approximate dynamics that are used in computing the BIF respectively. Hence, $\tilde\kappa$ takes the role of $\tilde p$ used earlier in our description. 
}  In that case $\forw{\kappa}(m,\mu)$ need not be a probability measure, even if $\mu$ is. This motivates the following definition.

\begin{defn}\label{def:guidedproc}
For a \ho{guided process} with backward kernel $\tilde\kappa$ we have 
\[
\backw{\tilde\kappa}(h) = \left(m, \tilde\kappa h\right),\quad \text{where} \quad  m(x,y) = \frac{h(y) }{(\tilde\kappa h)(x)}.
\]
If $\varpi\ge 0$ and $\mu$ is a probability measure, then
\[
 \forw{\kappa}(m, \varpi \cdot \mu)(\!\dd y) = (\varpi w_\kappa(m, \mu)) \cdot \nu(\!\dd y) \]
with the \hb{weight} $w_\kappa(m, \mu)$ and probability measure $\nu$ defined by 
\begin{equation}\label{eq:wnu}
\begin{split}
\nu(\!\dd y) &=  w^{-1}_\kappa(m, \mu) \int  \frac{h(y) }{(\tilde \kappa h)(x)  }  \mu(\!\dd x) \kappa(x, \dd y) \quad \text{and}\\ 
 w_\kappa(m,\mu) &= \iint  m(x, y) \kappa(x,\dd y) \mu(\dd x) =  \int \frac{(\kappa h)(x)}{(\tilde \kappa h)(x)} \mu(\dd x).
 \end{split}
 \end{equation}
\end{defn}
Note this definition is consistent with our previous definition of $\forw{}$.

\newthought{Joint application of the backward- and forward maps} can be written as 
\[ F(\kappa, \tilde\kappa) = \langle \mathcal{F}_\kappa \mid \mathcal{B}_{\tilde\kappa}\rangle. \]
Two kernels $\kappa_1$ and $\kappa_2$ can be composed to $\kappa_1\kappa_2$ and applied in parallel as $\kappa_1 \otimes \kappa_2$. \sidenote{Composition of Markov kernels follows from the Chapman-Kolmogorov equations: $ (\kappa_1\kappa_2)(x, \dd y) = \int \kappa_1(x, \dd z) \kappa_2(z,\dd y)$.} It turns out that
\begin{equation}\label{eq:optic}
\begin{split}
F(\kappa_1\kappa_2, \tilde\kappa_1\tilde\kappa_2) &= F(\kappa_1, \tilde\kappa_1)\cdot F(\kappa_2, \tilde\kappa_2)  	\\
F(\kappa_1\otimes\kappa_2, \tilde\kappa_1\otimes\tilde\kappa_2) &= F(\kappa_1, \tilde\kappa_1)\otimes F(\kappa_2, \tilde\kappa_2)  
\end{split}	
\end{equation}
I haven't told you about $\cdot$ and $\otimes$ on the right-hand-side. That is in the paper! Also be careful with interpreting $\otimes$: while we use the same symbol on the left- and right-hand-side, in the former case it is parallel application of Markov kernels but in the latter case denoting product measure.

\newthought{This is the beginning of a story} where the forward evolution of the Markovian process on the DAG is written as parallel/serial composition of Markov kernels. To each forward kernel we specify a backward kernel $\tilde\kappa$, which by the way \ho{need not necessarily be Markov}. Then each element $\kappa$ in this composition gets replaced with $F(\kappa, \tilde\kappa)$ in \hb{ABFFG}. That's it. Hence: ``all'' that needs to be implemented is
\begin{enumerate}
	\item the forward and backward map;
	\item the compositionality rules appearing in Equation \eqref{eq:optic}.
\end{enumerate}
Of course, we additionally need a dictionary which tells us in which order to compose in the forward evolution. 

\newthought{If you are familiar with reverse-mode automatic differentiation} (AD) you may have noted similarities. Indeed, the compositional structure here is essentially the category of \hb{optics} proposed for AD.


\section{Continuous time transitions over an edge}\label{sec:cont}

In many settings, the natural modelling framework is to assume that  the transition over an edge is in fact the result of evolving a continuous time process over some time interval. \sidenote{Phylogenetics is one example, where a Brownian motion or finite-state continuous time Markov process pops up.} 
Thus suppose along an edge the transition is the result of running a continuous-time Markov process $X$ over the interval $[0,T]$.
Conditioning the process on its value at time $T$ corresponds to a \hb{change of measure}, details follow shortly.  We closely  follow the exposition in the paper by Palmowski and Rolski from 2002 \cite{PalmowskiRolski2002}, which we denote {\sc PR2002}.

\newthought{Warning:} this section is necessarily mathematically more demanding, as continuous-time Markov processes are inherently  more complicated than their discrete-time counterpart.

\newthought{Assume $X_t$ is Markov process} on a filtered probability space  $(\Omega, \scr{F},\{\scr{F}_t\}, \PP)$ having extended generator $\scr{L}$ with domain $\scr{D}(\scr{L})$. \footnote{Recall that a Markov process is (under certain technical conditions) characterised by its infinitesimal generator $\scr{L}_t$. i.e.\ 
	\[ (\scr{L}_t f)(x) = \lim_{h\downarrow 0} t^{-1} \E \left[ f(X_{t+h}) - f(X_t) \mid X_t = x\right], \]
	for all $f$ in the domain of $\scr{L}$ (which is part of the definition and defined by those $f$ for which the above limit exists). }
	For a strictly positive function $f$  define 
\[ E^f(t) = \frac{f(X_t)}{f(X_0)} \exp\left(-\int_0^t \frac{(\scr{L} f)(X_s)}{f(X_s)} \dd s \right). \]
If $h$ is such that $E^h(t)$ is a martingale, then it is called an exponential martingale and then $h$ is called a \ho{good} function. As $\EE E^h(t) = \EE E^h(0)=1$ this martingale can be used to define a change of measure. 

Under this change of measure, the process $X_t$ is typically again Markovian with \textit{nicer} properties.\sidenote{The key example of ``nicer'' for us is that the process is conditioned on a future event.}  For a probability measure $\PP$ we denote its restriction to $\scr{F}_t$ by $\PP_t$.  The  main result of \textsc{PR2002} (Theorem 4.2) says the following:   if $h$ is a good function and  $\bar{\PP}_t$ is defined by 
\[ \frac{\dd \bar{\PP}_t}{\dd \PP_t} = E^h(t) \]
then under $\bar{\PP}_t$ the process $X_t$ is a Markov process with extended generator
\begin{equation}\label{eq:htransformedL} \bar{\scr{L}} f = \frac1{h} \left[ \scr{L}(f h) - f \scr{L} h\right]. \end{equation}
Moreover, $\scr{D}(\scr{L})=\scr{D}(\bar{\scr{L}})$.    Note that if $h$ is harmonic, i.e.\ $\scr{L} h=0$, then we have the simple expression $E^h(t) = h(X_t)/h(X_0)$.  

\newthought{How do we know $h$  is a good function} (meaning that  $E^h(t)$ is a martingale)? First, if we define 
	\[ D^f(t) = f(X_t) - \int_0^t \scr{L}(X_s) \dd s \]
then by Lemma 3.1 in \textsc{PR2002}, $\{D^f(t),\, t\ge 0\}$ is a local martingale if and only if 	$\{E^f(t),\, t\ge 0\}$ is a local martingale.\sidenote{This requires $f \in \scr{D}(\scr{L})$ but additionally $f$ needs to satisfy integrability conditions. We refer to the paper for details.}	 The \textit{local} martingale can be strengthened to true martingale under certain extra conditions on $h$ (sufficient conditions are given in Proposition 3.2 in \textsc{PR2002}).

\newthought{Now it is time to apply these results.} To this end, 
	we will apply the change-of-measure to the space-time-process $(t, X_t)$, which has infinitesimal generator $\scr{A} = \partial_t + \scr{L}$.
To condition the process $X$ on $\{X_T=x_T\}$ we take the specific choice \[ h(t,x) = p(t,x; T,x_T),\] where $p$ denotes the transition density of $X$, evolving from $x$ at time $t$ to $x_T$ at time $T$. 
	It is well known that for this choice of $h$ we have $\scr{A} h=0$, which is simply \ho{Kolmogorov's backward equation}.\sidenote{Put differently, $(t,x) \mapsto h(t,x)$ is space-time harmonic.} Define the measure $\PP^\star_t$ by 
\[ \frac{\dd \PP^\star_t}{\dd \PP_t} (X) = E^h(t)=\frac{h(t,X_t)}{h(0, X_0)}. \]
Using \eqref{eq:htransformedL} we can find the extended generator under $\PP^\star_t$ to be
\begin{equation}\label{eq:Lstar} \scr{L}^\star f = \frac1{h} \left[ \scr{L}(f h) - f \scr{L} h\right],\end{equation}
where $f$ depends on $(t,x)$. Carrying out this computation in concrete examples reveals for example that 
\begin{itemize}
	\item if $X_t$ is a diffusion process, then $X^\star_t$ is also a diffusion process with an extra term added to the drift parameter;
	\item if $X_t$ is a Poisson process of constant intensity, then $X^\star_t$ is a non-homogeneous Poisson process. 
\end{itemize}
	What does the change of measure imply? I claim that under $\PP^\star$ the process $X$ is conditioned on the event $\{X_T=x_T\}$. To see this, take $t_0 < t_1 < \cdots < t_n < t < T$, assume the process is started at $x_0$  and consider 
\begin{fullwidth}	
\begin{align*} & \EE^\star \left[\psi(X_{t_1}, \ldots, X_{t_n}, X_t)\right]  = \EE  \left[ E^h(t) \psi(X_{t_1}, \ldots, X_{t_n}, X_t) \right] \\ & \qquad = \int g(x_1,\ldots, x_n, x_t) \frac{p(t,x_t; T,x_T)}{p(t_0, x_0; T, x_T)} p(t_i, x_i; t, x_t)\prod_{i=1}^n p(t_{i-1}, x_{i-1}, t_i, x_i) \dd x_1 \cdots \dd x_n \dd x_t  \\ & \qquad =  \EE \left[\psi(X_{t_1}, \ldots, X_{t_n}, X_t) \mid X_T = x_T\right] \end{align*}
\end{fullwidth}
So far so good, but the problem is of course that only in very specific cases the transition densities $p$ are known. Therefore, in general $h$ is unknown and at first sight the preceding does not seem to be of any help. 	
However, suppose that there is a Markov process $\tilde X_t$ with space-time generator $\tilde{\scr{A}}$ and  tractable $g$ satisfying $\tilde{\scr{A}} g=0$. 
Let the measure $\PP^\circ_t$ be defined by
\[ \frac{\dd \PP^\circ_t}{\dd \PP_t}(X) = E^{g}(t). \]
Therefore
\[ \frac{\dd \PP^\star_t}{\dd \PP^\circ_t}(X) = \frac{h(t,X_t)}{h(0, X_0)} \frac{g(0, X_0)}{g(t,X_t)}  \exp\left(\int_0^t \frac{(\scr{A} g)(s,X_s)}{g(s,X_s)} \dd s \right). \]
The term in the exponential can be simplified slightly since we  have
\[ \scr{A} f = (\partial t + \scr{L}) f  = (\scr{L} - \tilde{\scr{L}}) f. \]
Again, using \eqref{eq:htransformedL} we can find the extended generator under $\PP^\circ_t$ to be
\begin{equation}\label{eq:generator_circ} \scr{L}^\circ f = \frac1{h} \left[ \scr{L}(f h) - f \scr{L} h\right]. \end{equation}

\newthought{Let's summarise some of our findings.} By a change of measure from $\PP$ to $\PP^\star$ the Markov process $X_t$ can be conditioned. Moreover, the expression for $\scr{L}^\star$ in \eqref{eq:Lstar} can be used to identify the dynamics of the process under $\PP^\star$. 

 Unfortunately, the $h$-function required for $\PP^\star$ is usually intractable and hence we take an approximation $g$ to $h$. This $g$ can be used in the same way for an exponential change of measure to define $\PP^\circ$. The process $X_t$ under $\PP^\circ$ is tractable and its dynamics can be identified using \eqref{eq:generator_circ}.  Finally, the likelihood ratio $\dd \PP^\star / \dd \PP^\circ$ is known in closed form. This quantity can be used to correct for  the discrepancy between $\PP^\star$ and $\PP^\circ$ in Monte-Carlo methods such as importance sampling, sequential Monte Carlo and Markov Chain Monte-Carlo. \sidenote[][-0.5in]{Note that the annotation by $\circ$ and $\star$ is consistent with our earlier use of these symbols for transition densities.}

\newthought{This is the basic idea.} There are definitely subtle things that need to be taken care of: most importantly, we need to assess the behaviour of the likelihood ratio as $t\uparrow T$. For certain classes of Markov processes, under certain extra conditions, it can be shown  that \sidenote{This is analogous to the expression in Equation \eqref{eq:lr}.}
\[ \frac{\dd \PP^\star_T}{\dd \PP^\circ_T}(X) = \frac{g(0,X_0)}{h(0, X_0)}   \exp\left(\int_0^T \frac{(\scr{A} g)(s,X_s)}{g(s,X_s)} \dd s \right). \]
Only $h(0, X_0)$ is still intractable, but as it shows up as a multiplicative constant in the denominator that turns out to be harmless. Note the similarity of this expression to \eqref{eq:lr}.


\section{Example: Stochastic Differential Equations on a tree}\label{sec:sde_tree}

To conclude, let's consider a toy example with an SDE on a directed tree. 
As the transition densities of the process are intractable, we adopt the approach we have outlined:
\begin{enumerate}
\item on each of the edges we define a function $g$;
\item the process $X^\circ$ is defined by applying Doob's $h$-transform using $g$.
\end{enumerate}
This means that on segments where the process evolves as a diffusion process, the process $X^\circ$, characterised by its extend generator $\mathcal{L}^\circ$ as specified in \eqref{eq:generator_circ}, is run forward. 
For discrete-transitions, the process evolves according to the transition densities $p^\circ$ as specified in \eqref{eq:circ}. 

What $g$ to use? It should be tractable and this tractability should be preserved in the backward filtering steps, starting from the leaves back to the root vertex. I'll illustrate with the
 setting depicted in  Figure \ref{fig:tree2}.
\begin{marginfigure}
\begin{center}
\begin{tikzpicture}

\tikzstyle{empty}=[fill=white, draw=black, shape=circle,inner sep=1pt, line width=0.7pt]
\tikzstyle{solid}=[fill=black, draw=black, shape=circle,inner sep=1pt,line width=0.7pt]

\begin{pgfonlayer}{nodelayer}
				\node [style=solid,label={$s$},] (2b) at (-6, 3.5) {};
		\node [style=solid,label={$t$},] (0) at (-4, 3.5) {};

		\node [style=empty,label={$v_1$},] (2a) at (-2, 4.5) {};

		\node [style=empty,label={$v_2$}] (2) at (-2, 2.5) {};
\end{pgfonlayer}

\begin{pgfonlayer}{edgelayer}
		\draw [style=edge] (2b) to (0);
		\draw [style=edge] (0) to (2);
		\draw [style=edge] (0) to (2a);
\end{pgfonlayer}

\end{tikzpicture}
\end{center}
	\caption{Part of a tree. On $s\to t$ a continuous-time Markov process evolves. Observations are at leaf-vertices $v_1$ and $v_2$.}
	\label{fig:tree2}
\end{marginfigure}
If we assume $v_i \mid x_t \sim N(L_i x_t, \Sigma_i)$, then
\[ g_{t\to v_i}(x) = \phi(v_i; L_i x, \Sigma_i),\qquad i =1, 2. \]
Because of Gaussianity,  we can write
\begin{equation}\label{eq:cFH} g_{t\to v_i}(x) = \exp\left(-c_i + F_i^\T x -\frac12 x^\T H_i x \right) \end{equation}
for  triplets $(c_i, F_i, H_i)$, with $c_i$ scalar valued, $F_i$ vector valued and $H_i$ matrix valued. \sidenote{In fact, we have $c_i = -\log \phi(v_i; 0, \Sigma_i)$, $F_i = v_i^\T \Sigma^{-1}_i L_i$ and $H_i= L_i^\T \Sigma_i^{-1} L_i$.}
Now at vertex $t$ we have the \ho{fusion}-step yielding
\[ g_t(x) = \prod_{i=1}^2 g_{t\to v_i}(x) \]
which can be interpreted as collecting all messages at vertex $t$ from its children. Clearly, $g_t$ can be represented by the triplet
$(c_1 + c_2, F_1 + F_2, H_1 + H_2)$.
Now suppose that  the branch connecting vertices $s$ and $t$ represents the evolution of a continuous time process $X$ on the time-interval $[s,t]$.\sidenote[][-0.5in]{There is a slight abuse of notation here, as $s$ and $t$ both denote a vertex and time.}
On this  segment, we define $g$ by solving on $(s,t]$
\begin{equation}\label{eq:Cauchy} (\partial_u + \tilde{\mathcal{L}}_u) g =0, \qquad g(t,\cdot) =g_t(\cdot).  \end{equation}
where $\tilde{\mathcal{L}}_u f(x) = \sum_i \tilde{b}_i(u,x) \partial_i f(x) + \sum_{i,j} \tilde{a}_{i,j}(u) \partial_{i,j} f(x)$. This is the infinitesimal generator of the process evolving according to the \ho{linear} SDE
\begin{equation}\label{eq:auxiliary_sde} \dd \tilde X_t = (B(t) \tilde X_t + \beta(t)) \dd t + \tilde\sigma(t) \dd W_t \end{equation} where $\tilde{b}(u,x) = B(u) x + \beta(u)$ and $\tilde a= \tilde \sigma \tilde\sigma^\T$. 
Solving the partial differential equation in \eqref{eq:Cauchy} is known as the \ho{Cauchy problem}. With the specific choice of a linear SDE the nice thing is that since $\bar h_t$ is of the form \eqref{eq:cFH}, then for $u\in (s,t]$ we have $g(u,x) = \exp\left(c_u + F_u^\T x -\frac12 x^\T H_u x \right)$. Hence, the functional form of $g$, where it is represented by a triplet $(c,F,H)$ is preserved.   Moreover, 
\begin{equation}
  \label{eq:backwardODE}
    \begin{split}
        \dd H(u) &= \left( -B(u)^\T H(u) - H(u)B(u)+H(u)\tilde{a}(u)H(u) \right)\dd u,\\
        \dd F(u) &= \left( -B(u)^\T F(u) + H(u)\tilde{a}(u)F(u) + H(u)\beta(u) \right)\dd u,\\
        \dd c(u) &= \left( \beta(u)^\T F(u) + \frac{1}{2}F(u)^\T \tilde{a}(u)F(u) - \frac{1}{2}\trace\left(H(u)\tilde{a}(u)\right) \right)\dd u.
    \end{split}
  \end{equation}
see for instance \cite{mider2020continuousdiscrete}, Theorem 2.5. \sidenote[][0.5in]{This is just backward filtering of linear SDE: a problem which has been solved decades ago.} 
In this way, $g$ can be defined recursively on the whole tree, starting from the leaves all the way back towards the root. This constitutes the \hb{backwards filtering step}. The main computational work consists of solving the ODEs in \eqref{eq:backwardODE}. This operation scales quadratically in the dimension of the diffusion process. Improved scaling can be obtained in case of sparsity in $B$, $\beta$ and/or $\tilde\sigma$. 

\newthought{For the forward guiding step}, we start from the root and evolve the process on ``continuous-time'' segments under the law $\PP^\circ$. From \eqref{eq:generator_circ} we can identify that $X^\circ$ is a diffusion process satisfying the SDE
\[ \dd X^\circ_t = \left(b(t,X^\circ_t) + a(t,X^\circ_t) (F(t)-H(t) X^\circ_t)   \right) \dd t + \sigma(t,X^\circ_t) \dd W_t, \]
which can easily be forward simulated using Euler-discretisation (or more sophisticated SDE-solvers). 

\subsection{Numerical example using \texttt{MitosisStochasticDiffEq.jl}}\label{sec:num}

We illustrate the methods described in here using an example of an SDE on a tree. Frank Sch\"afer gave a $3$-minute talk about this a JuliaCon2021 \url{https://www.youtube.com/watch?v=rie7MTvPpIs}. The forward model is as follows: on each branch of the tree the process evolving according to the SDE \sidenote{The ``$.$'' appearing in the drift means that the $\tanh$ function is applied coordinatewise.}
\[ \dd X_t = \tanh.\left( \begin{bmatrix} -\th_1 & \th_1 \\ 
 \th_2 & -\th_2 \end{bmatrix} X_t \right) \dd t + \begin{bmatrix} \sigma_1 & 0 \\ 0 & \sigma_2 \end{bmatrix} \dd W_t . \]	
Now forward simulating from this model on given tree gives rise to the following paths:
\begin{fullwidth}
\begin{center}
\includegraphics[scale=0.5]{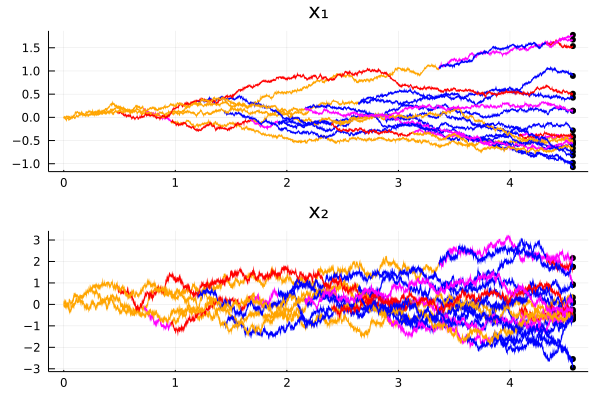}	
\end{center}
	\end{fullwidth}

We assume, as throughout, that only the values at the leaf-vertices are observed. Assume the tree-structure itself is known. We aim to estimate the parameters $\bs{\th}:=(\th_1, \th_2, \sigma_1, \sigma_2)$. Note that a standard Kalman-filter cannot be used due to the nonlinearity in the drift. We employ flat priors and use an MCMC-algorithm that iteratively updates the unobserved paths conditional on $\bs{\theta}$ and the observations, and $\bs{\theta}$ conditional on the unobserved paths. Elements of $\bs{\th}$ were updated using random-walk Metropolis-Hastings steps. The missing paths were updated using the BFFG-algorithm, where $\tilde{X}$ is chosen as in \eqref{eq:auxiliary_sde}, with \sidenote[][-1.0in]{We choose the diffusivity of $\tilde{X}$ to match that of $X$. This is crucial in case the extrinsic noise level approaches zero.}
\[ B = \begin{bmatrix} -\th_1 & \th_1 \\ 
 \th_2 & -\th_2 \end{bmatrix} \qquad \beta =\begin{bmatrix} 0 \\ 0 \end{bmatrix} \qquad \tilde\sigma = \begin{bmatrix} \sigma_1 & 0 \\ 0 & \sigma_2 \end{bmatrix}. \]
 \sidenote[][-1.0in]{The figures here are meant to illustrate the potential of the method. The code for producing the figures in this example is on the Github repository of the \texttt{MitosisStochasticDiffEq.jl}-package.}
Here are traceplots after running the algorithm for $10\_000$ iterations
\begin{fullwidth}	
\begin{center}
	\includegraphics[scale=0.5]{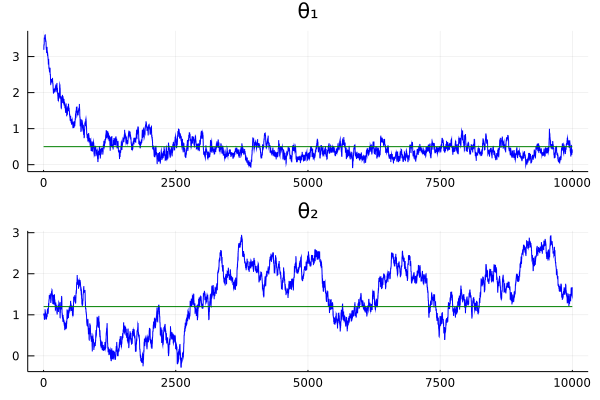}
\end{center}
\end{fullwidth}

\begin{fullwidth}	
\begin{center}
	\includegraphics[scale=0.5]{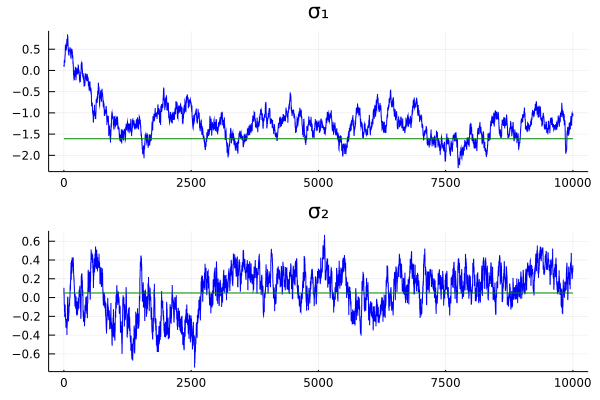}
\end{center}
\end{fullwidth}

One thing which makes this problem mildly difficult is that there is no strong nonlinearity in the drift. If that were the case, the paper {\it Continuous-discrete smoothing of diffusions}\cite{mider2020continuousdiscrete} gives a host of methods to deal with this setting, essentially choosing the process $\tilde X$ in a more advanced fashion. 

\section{Online talk}

I also tried to explain this in an online talk: \\ \url{https://www.youtube.com/watch?v=XjBO4GSc0i8}

\newthought{Acknowledgement:} Thanks to \hb{Frank Sch\"afer} (University of Basel) and \hb{Stefan Sommer} (University of Copenhagen) for providing detailed feedback on earlier versions that helped improving this paper.

%

\bibliography{bffg_simple2.bib}
\bibliographystyle{plainnat}

\end{document}